\documentclass[chicago]{emulateapj} \newif\ifdraft \drafttrue   \newif\ifpasp \draftfalse
\usepackage{graphicx}
\usepackage{natbib}
\usepackage{color}
\usepackage[linktocpage=true]{hyperref}
\hypersetup{
            colorlinks=true,
            linkcolor=black,
            filecolor=black,
            citecolor=black,
            urlcolor=blue
           }

\newcommand{\getlength}[1]{\ifx#1\end \let\next=\relax
            \else\advance\count255 by1 \let\next=\getlength\fi \next}

\newcommand{\ifnularg}[1]{ \count255=0 \getlength#1\end \ifnum\count255=0 }
\newcommand{\ifm}{\makebox{}\ifmmode}
\long\def\ifundefined#1#2#3{\expandafter\ifx\csname
  #1\endcsname\relax#2\else#3\fi}
\newcommand{\beq}   { \begin{eqnarray} }
\newcommand{\eeq}[1]{ \ifnularg{#1} end{eanarray} \else
                      \label{#1}\end{eqnarray}    \fi }
\newcommand{\eeql}   { \end{eqnarray} }
\newcommand{\eeqn}   { \nonumber \end{eqnarray} }

\newcommand{\ceduna}{\mbox{\sc ceduna}}
\newcommand{\Wark}{\protect \mbox{\sc wark\hspace{0.02em}{\footnotesize{1\hspace{-0.01em}2}}\hspace{-0.0em}m}}
\newcommand{\wark}{\protect \mbox{\sc wark\hspace{0.02em}{\footnotesize{3\hspace{-0.01em}0}}\hspace{-0.0em}m}}
\newcommand{\hobart}{\mbox{\sc hobart\hspace{0.04em}{\footnotesize{2\hspace{-0.01em}6}}}}
\newcommand{\Fermi}{{\it Fermi}}
\newcommand{\PIMA}{$\cal P\hspace{-0.067em}I\hspace{-0.067em}M\hspace{-0.067em}A$ }

\definecolor{Dred}{rgb}{0.312,0.070,0.070}
\definecolor{Dblue}{rgb}{0.070,0.070,0.312}
\definecolor{Dgreen}{rgb}{0.070,0.312,0.070}
\definecolor{Db}{rgb}    {0.050,0.0,0.320}

\newcommand{\Blb}[1]{\textcolor{Dblue}{\bf #1}}

\newcommand{\web}[1]{\Blb{\url{#1}}}

\newcommand{\timezone}{-0400}
\newcommand{\Number}[1]{\ifnum#1<10\relax0\number#1\else\number#1\fi}
\newcommand{\isodate}{
\count151=\time
\divide\count151 by 60
\count151=\count151
\multiply\count151 by 60
\count152=\time
\advance\count152 by -\count151
\divide\count151 by 60
\count152=\count151
\multiply\count151 by 60
\count153=\time
\advance\count153 by -\count151
\Number{\year}.\Number{\month}.\Number{\day}--\Number{\count152}:\Number{\count153} \enskip \timezone
}

\normalfont

\newcounter{note}
\let\oldmarginpar\marginpar
\renewcommand\marginpar[1]{\-\oldmarginpar[\raggedleft\footnotesize #1]%
{\raggedright\footnotesize #1}}
\newcommand{\Note}[1]{\Rdb{#1}{\addtocounter{note}{1}%
\marginpar{\small\underline{\Rdb{Corr \arabic{note}}}}}}

 \renewcommand{\Note}[1]{#1}
\ifdraft\voffset=-20mm\fi
\ifdraft\renewcommand\submitted{Received March 23, 2015;Accepted April 8, 2015}\fi
\renewcommand\submitted{Received March 23, 2015;Accepted April 8, 2015}

\begin{document}

\shorttitle{First scientific VLBI observations with WARK30M}
\shortauthors{Petrov et al.}
\title{First scientific VLBI observations using New Zealand 30 meter radio telescope WARK30M}

\author{L. Petrov}
\affil{Astrogeo Center, Falls Church, VA 22043, USA}
\email{Leonid.Petrov@lpetrov.net}

\author{T. Natusch, S. Weston}
\affil{Institute for Radio Astronomy and Space Research, 
       Auckland University of Technology, Private Bag 92006,
       Auckland 1142, New Zealand}

\author{J. McCallum, S. Ellingsen}
\affil{University of Tasmania, Private Bag 37, Hobart TAS 7001, Australia}

\and
\author{S. Gulyaev}
\affil{Institute for Radio Astronomy and Space Research, 
       Auckland University of Technology, Private Bag 92006,
       Auckland 1142, New Zealand}

\begin{abstract}
%
%We report the results of a successful 24 hour 6.7 GHz VLBI experiment using the 30 meter radio telescope WARK30M near Warkworth, New Zealand, recently converted from a radio telecommunications antenna, and two radio telescopes located in Australia: Hobart 26-m and Ceduna 30-m. The geocentric position of WARK30M is is determined with a 100 mm uncertainty  for the vertical component and 10 mm for the horizontal components. We report correlated flux densities at 6.7 GHz of 175 radio sources associated with Fermi gamma-ray sources. A parsec scale emission from the radio source 1031-837 is detected, and its association with the gamma-ray object 2FGL J1032.9-8401 is established with a high likelihood ratio. We conclude that the new Pacific area radio telescope WARK30M ready to operate for scientific projects.
%

We report the results of a successful 24~hour 6.7~GHz VLBI experiment using
the 30 meter radio telescope \wark\ near Warkworth, New Zealand, recently 
converted from a radio telecommunications antenna, and two radio telescopes 
located in Australia: Hobart 26-m and Ceduna 30-m. The geocentric position 
of \wark\ is determined with a 100~mm uncertainty  for the vertical component 
and 10~mm for the horizontal components. We report correlated flux densities 
at 6.7 GHz of 175 radio sources associated with \Fermi\  $\gamma$-ray sources. 
A parsec scale emission from the radio source 1031-837 is detected, and its 
association with the $\gamma$-ray object 2FGL J1032.9-8401 is established with 
a high likelihood ratio. We conclude that the new Pacific area radio telescope 
\wark\  is ready to operate for scientific projects.

\end{abstract}
\keywords{
          Astronomical Instrumentation
}

\maketitle

\section{Introduction} \label{s:intro}
\ifdraft \par\vspace{-120ex} {\large\it\hspace{-5mm} Draft of \isodate} \par \vspace{117ex}\par \fi

The New Zealand 30-m radio telescope facility is located near the township 
of Warkworth, New Zealand, just 5 km from the Pacific coast. It was built 
in 1984 as a telecommunication antenna (the Earth Station) to be used by the 
New Zealand Post Office, later by Telecom New Zealand. In 2010 it was 
transferred to the Institute for Radio Astronomy and Space Research (IRASR) 
of Auckland University of Technology for conversion to a radio telescope.  

\begin{figure*}
   \begin{center}
      \ifpasp
          \includegraphics[width=0.85\textwidth]{fig1.eps}
        \else
          \includegraphics[width=0.80\textwidth]{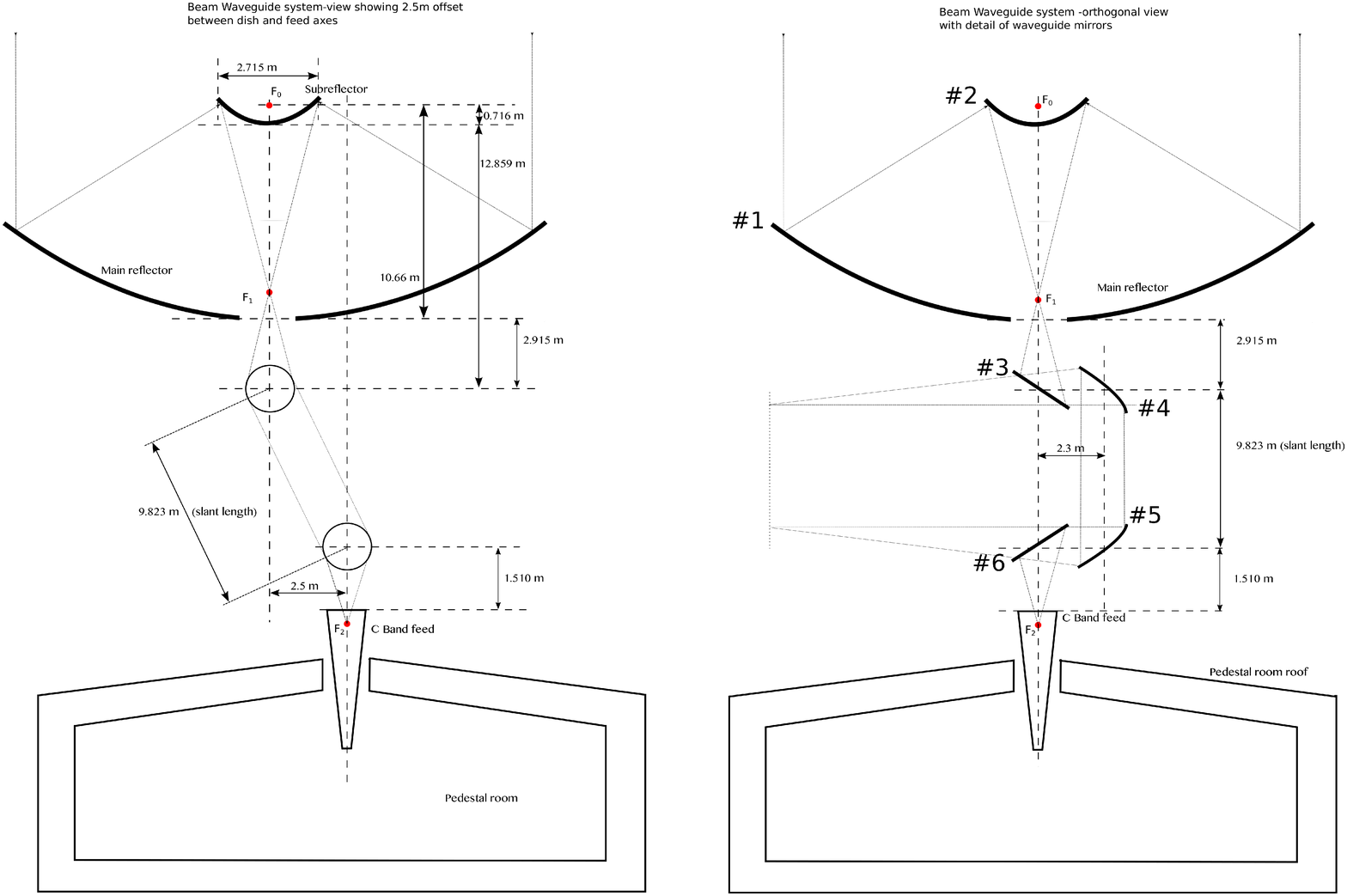}
      \fi
   \end{center}
   \caption{{\sc wark30m} beam wave-guide diagram in two orthogonal 
            projections (not to scale). Radio waves from the sky are reflected 
            from the main mirror \#1, then reflected from the secondary mirror
            and guided to the diagonal mirror \#3, which has a circular shape 
            in the plane perpendicular to the elevation axis. The parallel 
            beam formed by this mirror travels 2.3~m towards the elliptical 
            mirror \#4 that reflects the beam into the waveguide that is 
            tilted at $14.74^\circ$ with respect to the vertical. The 
            elliptical mirror \#5 sends the beam to the diagonal mirror \#6, 
            which directs it into the feed horn mounted in the antenna 
            pedestal.
   }
   \label{f:bwg}
   \par\vspace{2ex}
\end{figure*}

The 30-m antenna has a special wheel-on-track beam-waveguide (BWG) design. 
This design is often used for deep space network and telecommunications 
\citep{r:bwg}. The feed horn and the receiver are located in the 
pedestal and do not rotate when the antenna moves along the azimuth and
elevation axes. A system of six mirrors directs the beam into the 
receiver (see Figure~\ref{f:bwg}). The design of the \wark\ antenna 
should be taken into account for data reduction when dual-polarized 
data are processed.

  Conversion of the 30-m antenna  took three years and required a number of
upgrades and repairs. Among them was enabling the telescope to track sources
with pointing accuracy $1'$, connecting the telescope with a 10~Gbps Internet,
installation of an uncooled  C-band receiver that was donated to IRASR 
by Jodrell Bank Observatory, installation of a new Digital Base Band Converter 
(DBBC-2) \citep{r:dbbc2}, installation of a Mark5B+ VLBI recorder, connection 
to the observatories existing Hydrogen maser via a Symmetricom fibre-optic 
distribution amplifier. Detailed description of the upgrades and the 
specification of the Warkworth 30-m radio telescope can by found in 
\citet{r:woodburn}. This work was completed in late 2014, fringe checks 
were performed and the antenna was ready for its first scientific VLBI 
experiment.

  As a part of commissioning new antennas, the position of the antenna 
reference point should be determined. A reference point is defined as the 
point of the projection of the movable elevation axis onto the fixed 
azimuthal axis. \Note{For the analysis of VLBI source imaging experiments 
made in a phase-referencing mode, position of VLBI stations should 
be known with accuracy 10--20~centimeters to avoid noticeable image 
smearing\footnote{See \web{http://www.evlbi.org/user_guide/phrefmp.html}
for an example of the effect of position errors in quality of image
restoration}~\citep[see e.g.][]{r:cha2002}).} For astrometry 
applications, angular position accuracies of tens of $\mu$as are required, 
making the requirements on the accuracy of station positions much more 
stringent: 5--10~mm. One way of estimating the position of the antenna 
reference points is through analysis of a combination of a ground survey 
of markers attached to the antennas from a local network around the station 
and GPS observations from the points in the local network 
\citep[see e.g.,][]{r:sarti04,r:sarti09}. At the time of writing this had 
been planned for the Warkworth 30~m antenna but not carried out. 
An alternative procedure for estimation of station position is to use 
the VLBI technique to determine group delays and then derive the reference 
point position from these group delay measurements. The advantage of the 
latter approach is that it also provides useful diagnostics of the VLBI 
equipment and can be combined with astronomical tasks.

  Here we report results from the first scientific experiment  
conducted with \wark\ and two Australian radio telescopes operated by
the University of Tasmania in Hobart and Ceduna \citep{r:mcc05} on 
December 11, 2014. 
The VLBI experiment settings are presented in 
section \ref{s:exp}. The data analysis for both geodetic and 
astronomical purposes is described in section  \ref{s:anal}. 
Concluding remarks are made in section \ref{s:concl}.

\section{VLBI Experiment}  \label{s:exp}

\begin{figure}[h]
   \ifpasp
        \includegraphics[width=0.48\textwidth]{fig2.eps}
      \else
        \includegraphics[width=0.48\textwidth]{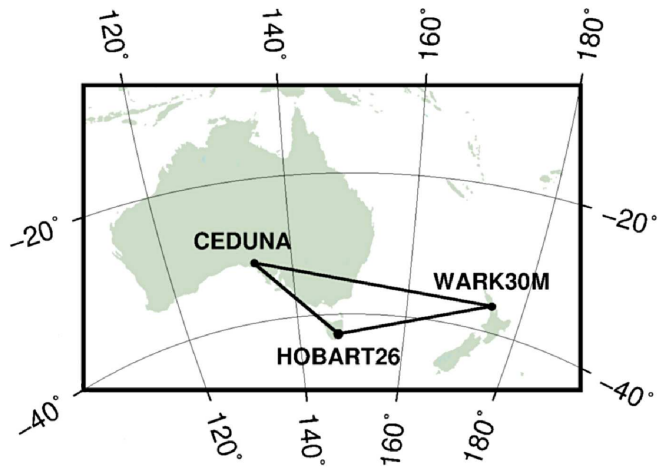}
   \fi
   \par\vspace{-1ex}\par
   \caption{VLBI network of this experiment. The map shows locations 
            of the three radio telescopes in New Zealand and Australia
            and the corresponding baselines.}
   \label{f:map}
\end{figure}

  Since the \wark\ C-band receiver has a relatively narrow band,
approximately 300~MHz, our choice of frequency range for VLBI
observations was rather limited. We recorded right-circular polarization 
in a range [6.592, 6.848]~GHz. The signal was split into 16 intermediate
frequencies (IFs) of 16~MHz each and recorded with 2 bit sampling and 
an aggregate bit rate 1~Gbps. This setup allows us to determine group 
delay with formal uncertainty 215~ps when the signal to noise ratio (SNR) 
is 10. Two radiotelescopes in Australia participated: \ceduna\ and 
\hobart\ (Figure~\ref{f:map}).

  We set three goals for the first scientific experiment:

\begin{itemize}
   \item to determine the position of \wark\ with centimeter level 
         of accuracy;

   \item to test the effectiveness of the southern VLBI array with 
         \wark\ for monitoring radio flux density of $\gamma$-ray 
         sources;

   \item to test the usability of the southern VLBI array with 
         \wark\ for association of $\gamma$-ray sources discovered 
         with \Fermi\ by detecting their parsec-scale radio emission.
\end{itemize}

  The pool of 483 targets included all sources from the Radio Fundamental
Catalogue\footnote{The catalogue is available at \web{http://astrogeo.org/rfc}} 
(Petrov and Kovalev, 2015, in preparation) that are known to be brighter than
200~mJy at 8~GHz on baselines longer than 5000~km and are associated with 
$\gamma$-ray sources. At the moment, 20\% of sources with parsec-scale 
emission brighter than 200~mJy are known to be $\gamma$-ray loud. Restricting 
the source list only to these objects did not affect our ability to measure 
station positions. In total, 244 target sources were scheduled.
This includes 7 radio sources from a dedicated Australia Telescope Compact 
Array (ATCA) 5.5/9.0~GHz survey \citep{r:aofus2} never before observed with 
VLBI. These sources are located within the error ellipse of \Fermi\ objects 
but marked as ``unassociated'' in the the 2FGL catalogue \citep{r:2fgl}. As 
shown by \citet{r:aofus2}, detection of parsec-scale radio emission is 
a powerful method for association of $\gamma$-ray sources. Since the number 
of compact radio sources is limited, as a rule of thumb, the probability of
finding a compact background source with flux density from parsec scales
brighter than 10~mJy at 8~GHz within the error ellipse of a $\gamma$-ray 
object is less than 10\%. 

  Observations were scheduled with software 
{\sf sur\_sked}\footnote{See \web{http://astrogeo.org/sur\_sked}} in 
a sequence that minimizes slewing time and complies with the constraints 
that (a)~the next source is located at a distance not less than $20^\circ$ 
and (b)~the minimum temporal separation between consecutive observations 
of a target source is 3~hours. Target sources were observed for 90~s each, 
except for those seven sources in the vicinity of unassociated 2FGL objects: 
they were observed for 300~s each. Every hour a sequence of 4 scans of 
atmosphere calibrators was inserted in the schedule: two scans of sources 
at elevation angles in a range $[15^\circ, 30^\circ]$ and two sources at 
elevation angles in a range $[60^\circ, 88^\circ]$. The purpose of including 
these scans in the schedule was to improve the robustness of estimation 
of residual atmospheric path delay in the zenith direction.

\section{Data Analysis}    \label{s:anal}

\begin{figure*}
   \ifpasp
        \includegraphics[width=0.455\textwidth]{fig3a.eps}
      \else
        \includegraphics[width=0.455\textwidth]{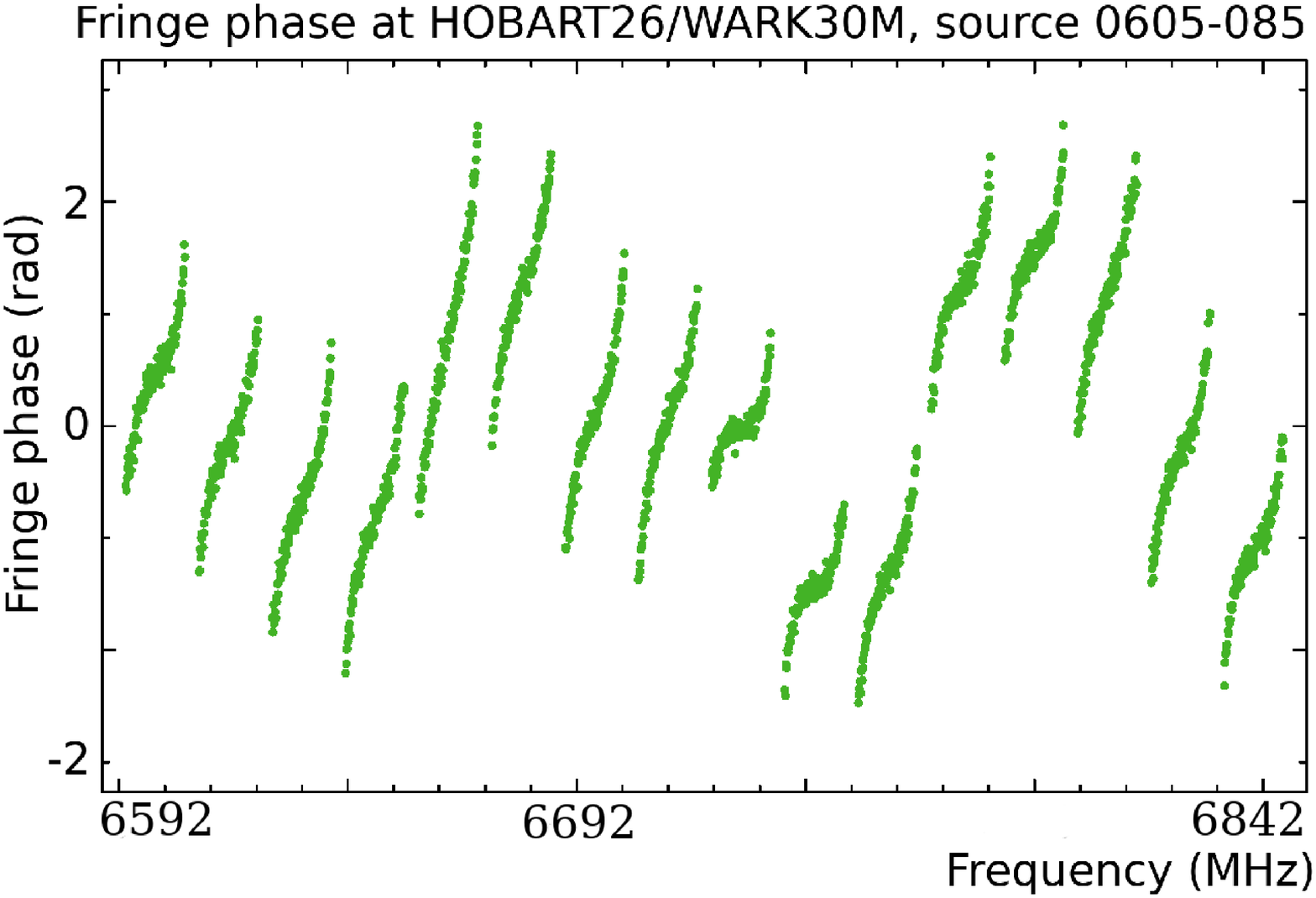}
   \fi
   \hspace{0.064\textwidth}
   \ifpasp
        \includegraphics[width=0.48\textwidth]{fig3b.eps}
      \else
        \includegraphics[width=0.48\textwidth]{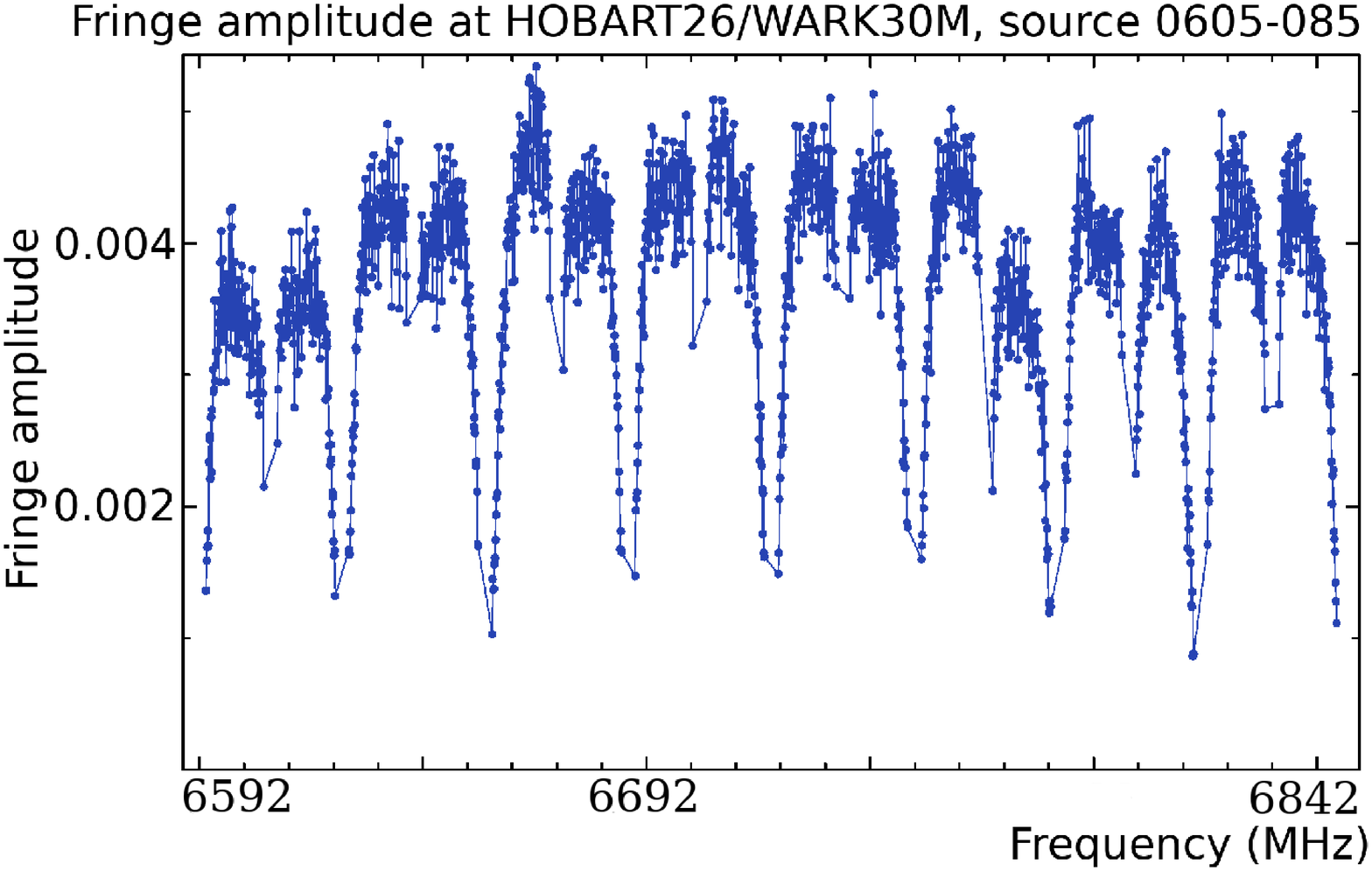}
   \fi
   \caption{Typical fringe phase ({\it Left}) and fringe amplitude
            ({\it Right}) at {\sc hobart26/wark30m} baseline before
            applying the complex bandpass.}
   \label{f:fp}
   \ifdraft \par\vspace{-1ex}\par \fi
\end{figure*}

\begin{figure}[h]
   \ifpasp
        \includegraphics[width=0.46\textwidth]{fig4.eps}
      \else
        \includegraphics[width=0.46\textwidth]{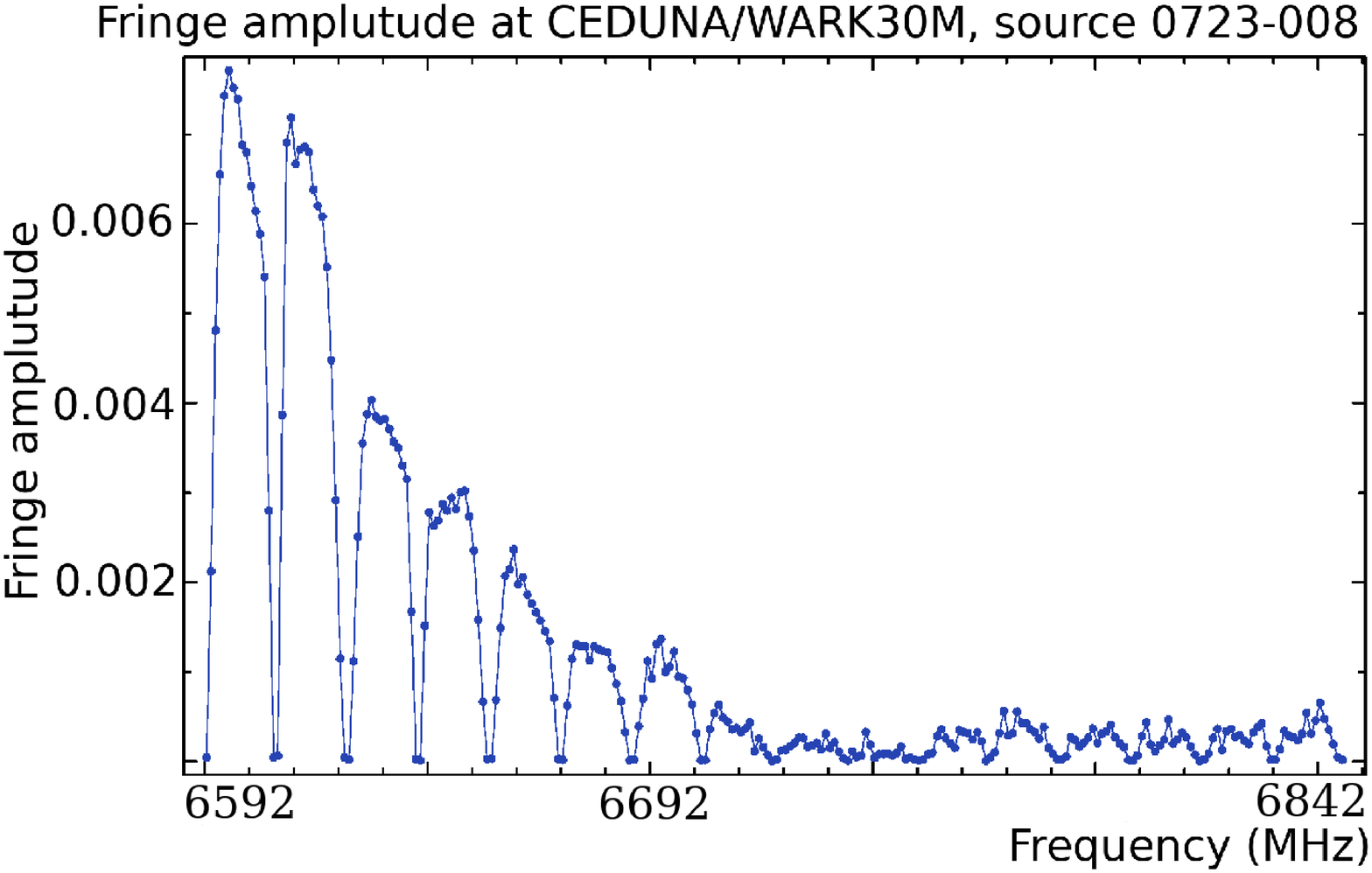}
   \fi
   \caption{Fringe amplitude {\sc ceduna/wark30m} baseline without 
            applying the complex bandpass. Receiver bands at 
            these two stations do not match.} 
   \label{f:cdwa}
   \ifdraft \par\vspace{-1ex}\par \fi
\end{figure}

  The data were transferred electronically to Hobart and correlated with 
the DiFX software \citep{r:del07}. Further processing was undertaken with 
the software package \PIMA\footnote{See \web{http://astrogeo.org/web}}. 
We determined the complex bandpass using several strong sources. After 
applying the bandpass calibration, we performed fringe fitting for each 
observation, i.e. we determined group delay, phase delay rate, and group 
delay rate that maximizes the coherent sum of visibilities over time and 
frequency \citep{r:vgaps}. The fringe plots for the \hobart/\wark\ baseline 
were as expected with no obvious anomalies (See an example in 
Figure~\ref{f:fp}). Unfortunately, the receiver frequency response at 
\ceduna\ did not match the response from other telescope receivers. Fringe 
amplitude for baselines to \ceduna\ at frequencies higher than \Note{6.620~GHz}
dropped by a factor of 10 (Figure~\ref{f:cdwa}). 

The SNR for baselines to \ceduna\ was typically a factor 4 worse than 
expected, and, therefore, uncertainty in group delays were higher by the 
same factor. This resulted in the non-detections of 19\% of target sources 
due to the corresponding sensitivity degradation. Nevertheless, the 
experiment did not fail, although position accuracy achieved was lower 
than expected.

\subsection{Geodetic Analysis}    \label{s:geodesy}

  Further processing was made with VLBI analysis software 
VTD/Post-Solve\footnote{See \web{http://astrogeo.org/vtd}}. After evaluation 
of group delays from visibility data, the theoretical path delays were 
computed using the state-of-the art model and small differences between them 
and the measured group delays were formed. A detailed description of this 
step can be found in \citet{r:awark}.

  The a~priori slant path delays in the neutral atmosphere in the direction 
of observed sources were computed through numerical integration of differential 
equations representing the wave propagation through the heterogeneous media. 
The four-dimensional field of the refractivity index distribution was 
computed using the atmospheric pressure, air temperature, and specific 
humidity taken from the output of the Modern Era Retrospective-Analysis for 
Research and Applications (MERRA) \citep{r:merra}. The output of the model 
presents the atmospheric parameters at a grid $0.67\degr \times 0.50\degr \times$
72 pressure levels $\times 6^h$ . A priori slant path delays in the ionosphere
were computed using global ionospheric maps of the total electron 
content derived from Global Navigation Satellite System observations by 
the analysis center CODE.

  In the initial least square (LSQ) solution, positions of all stations, 
except \hobart, were estimated, as well as the coefficients of the expansion 
for the  clock function and the residual atmosphere path delay in zenith 
direction into the B-spline basis of the 1st degree. During the initial phase 
of the data analysis, outliers were eliminated and the baseline-dependent 
corrections to the a~priori weights defined to be reciprocal to 
formal uncertainties of group delays were determined in such a way that
the ratio of the weighted sum of squares of residuals to their mathematical
expectation was close to unity. The weighted root mean squares (wrms) of
postfit residuals was 114~ps on \hobart/\wark\ baseline and 440~ps on
baselines to \ceduna. The increase of residuals with \ceduna\ was caused
by the mismatch of the receiver's frequency response.

  The final LSQ solution used all VLBI group delays collected from January 
1984 through January 2015, in total, 11.4 million values, including 
705 group delays from this experiment. Positions of all stations, 
coordinates of all sources, the Earth orientation parameters, as well as 
over 1 million nuisance parameters were estimated in a single LSQ run. 
The differences in velocity of \wark\ and \Wark\ was constrained to zero. 
Station \wark\ has a significant axis offset, the distance between the azimuth 
fixed axis and the moving elevation axis. We have adjusted that parameter
in our VLBI data analysis. Minimal constraints were imposed to require that 
the net-translation and net-rotation over new position estimates of 44 stations 
with long history with respect to positions of observations to these stations 
in the ITRF2000 catalogue \citep{r:itrf2000} be zero. This ensures that the
positions of all stations, including \wark\ be consistent with the 
ITRF2000 catalogue. More details about the parameter estimation technique can 
be found in \citet{r:rdv}. 

  The position estimate of \wark\ at epoch 2000.01.01 is provided in 
Table~\ref{t:coord}. The errors reported in the table are the formal 
uncertainties from the LSQ solution computed in accordance with the error 
propagation law. Projected to the local coordinate system, the uncertainties 
are 105, 14, and 12~mm for the Up, East, and North components respectively. 
High uncertainty in the Up component is due the necessity to estimate 
antenna axis offset. These two parameters correlate. If we choose not to 
estimate axis offset, but keep its value to 2500~mm, as specified by the 
antenna manufacturer, the position uncertainties drop to 38, 12, 
and 9~mm respectively.

\begin{table}[t]
   \caption{\sc Coordinates and velocities of {\sc wark30m} derived from analysis 
            of the VLBI experiment on epoch 2000.0. 
            \label{t:coord}
            }
   \par\medskip\par
   \begin{center}
      \begin{tabular}{l r @{$\;$} r @{$\;$} r l}
          \hline
                     X & $  -5115425.60 $ & $\pm$ & 0.08 & m \\
                     Y & $    477880.31 $ & $\pm$ & 0.02 & m \\
                     Z & $  -3767042.81 $ & $\pm$ & 0.06 & m \\
           Axis offset & $         2.61 $ & $\pm$ & 0.06 & m \\
             $\dot{X}$ & $       -19.6  $ & $\pm$ & 1.2  & mm/yr \\
             $\dot{Y}$ & $        -2.5  $ & $\pm$ & 0.5  & mm/yr \\
             $\dot{Z}$ & $        34.1  $ & $\pm$ & 1.0  & mm/yr \\
          \hline
      \end{tabular}
   \end{center}
   \par\medskip\par
   Since velocity of {\sc wark30m} was strongly constrained to {\sc wark12m}, 
   the reported velocity is based primarily on geodetic observations 
   of {\sc wark12m} station.
   % \par\vspace{-3ex}
\end{table}

  The largest source of systematic error is the path delay in the ionosphere.
In order to evaluate the contribution of the residual ionosphere to site 
position estimates, we re-processed the CONT-14 VLBI campaign with 
participation of \hobart\ and \Wark. During 14~days in May 2014 a~17-station 
array observed a geodetic schedule. We processed fourteen 24-hour 
experiments independently and estimated baseline lengths between \hobart\ 
and \Wark. We undertook two separate analyses. We used ionosphere-free 
linear combinations of group delays at 2.3 and 8.4~GHz in the first run. 
The so-called baseline length repeatability, i.e. the wrms of baseline 
lengths, was 2.7~mm. In the second run we used group delays at 8.4~GHz 
and applied the a~priori ionosphere contribution computed from the CODE 
global ionosphere model using the same algorithm as we used for analysis 
of the experiment with \wark. The wrms was 6.2~mm, i.e. the contribution 
of residual ionosphere to baseline lengths was 5.5~mm in quadrature. 
For deriving the contribution of the ionosphere to the vertical component 
of the \wark\ position estimate, we should scale the repeatability of the 
2415~km long baseline \hobart/\Wark\ by two factors: the geometric factor, 
the ratio of the Earth's diameter and the baseline length, and the frequency 
factor: the square of the effective frequency observed in CONT-14 campaign 
to the square of the effective frequency observed in our experiment. This 
gives us an estimate of the systematic error in the vertical component of 
\wark\ position: 50~mm. The systematic error in the horizontal component 
of position vector is smaller by the geometric factor: 10~mm. This analysis
demonstrates that the systematic errors due to the mismodeled ionosphere 
contribution do not dominate over random errors.

\subsection{Astronomy Analysis}    \label{s:astronomy}

   Among seven target sources in the vicinity of \Fermi\ objects
observed in our experiments, one radio source, {\sf 1031-837}, has been 
detected. \Note{Its coordinates, determined from the global LSQ 
solution}, are $ \alpha = 10^h30^m15^s.286 \pm 0^s.003, 
\delta = -84\degr03'08''.652 \pm 0''.004$ for the J2000.0 epoch. 
No $\cos\delta$ factor was applied to the reported uncertainty in
right ascension.

   At the time of our experiment, concurrent measurements of 
system temperature were not implemented. The system temperature at different 
elevations was measured eight days before the experiment, and we
found that it fits well to the following regression model:
$T_{\rm sys} = 82.6 + 4.3/\sin{E}$, where $E$ is the elevation angle
and the temperature is in Kelvins. A~priori measurements of the system
equivalent flux density (SEFD) resulted in the value of $\sim\! 600$~Jy.

   Since imaging the data of several scans from a three-element array
without concurrent $T_{\rm sys}$ measurements at one of the antennas is 
hopeless, we resorted to a simplified amplitude analysis that we have 
used before for fringe search surveys \citep{r:kcal,r:qcal}. Using the 
measured $T_{\rm sys}$ (a modelled $T_{\rm sys}$ in the case of \wark\ ) 
and the a~priori gain curves, we computed a~priori SEFD for each 
observation and calibrated fringe amplitudes averaged over time and 
frequency. In a case of \ceduna\ only the fringe amplitude from the 
first IF (6.592--6.608~GHz) was used. We identified 48 sources with 
known images from the VLBI Image Database\footnote{Available at 
\web{http://astrogeo.org/vlbi\_images}} maintained by the Astrogeo 
Center. Using these images, we computed the expected fringe amplitude 
at given baseline projections. We assumed that the image is the same 
at 8.4 and 6.7~GHz. We formed the ratios of the measured fringe 
amplitude to their expected values and solved for logarithms of 
antenna-based gain factors. Estimates of the SEFD at $45^\circ$ 
elevation angle are given in Table~\ref{t:sefd}. The anomalously high 
SEFD at {\sc hobart} is due to a failure of the cryogenic equipment
that resulted in a system temperature rise to $\sim\! 200$~K 
during the experiment. 

\begin{table}[h]
   \caption{\sc Average fitted SEFD at $45^\circ$ elevation angle.
            \label{t:sefd}
           }
   \par\vspace{-2ex}\par
   \begin{center}
       \begin{tabular}{lr}
          \ceduna    &  890 Jy \\
          \hobart    & 1890 Jy \\
          \wark      &  650 Jy \\
       \end{tabular}
   \end{center}
   The error of SEFD estimates is 15\%.
\end{table}

   Although the flux density of an individual source may change due 
to variability, the average ratio of the instantaneous flux density at
the experiment epoch to its value on the image epoch for the ensemble of 
48~sources should be stable. Using a relatively large ensemble for the
evaluation of gain factors allows us to estimate their statistical
uncertainty: 15\%.

\begin{table}[h]
   \caption{\sc The median correlated flux density
            at 6.7 ~GHz determined from analysis of this experiment.
            \label{t:flux}
   }
   \par\vspace{-2ex}\par
   \begin{center}
   \begin{tabular}{ll c @{\hspace{0.5em}} c @{\hspace{1.0em}} r @{\hspace{1.0em}} r}
       IAU Name   & IVS Name & (3) & (4) & (5) & (6) \\
       \hline
       \tt J0004-4736 & \tt 0002-478 & C &  6 & 0.38 & 0.03 \\
       \tt J0012-3954 & \tt 0010-401 &   &  3 & 0.36 & 0.03 \\
       \tt J0019+2021 & \tt 0017+200 & C &  3 & 1.03 & 0.04 \\
       \tt J0038-2459 & \tt 0035-252 &   &  3 & 0.71 & 0.03 \\
       \tt J0049-5738 & \tt 0047-579 &   &  3 & 0.68 & 0.03 \\
       \tt J0050-0929 & \tt 0048-097 & C &  3 & 0.36 & 0.03 \\
       \tt J0058-0539 & \tt 0055-059 & C &  3 & 0.35 & 0.02 \\
       \tt J0058-5659 & \tt 0056-572 &   &  3 & 0.44 & 0.03 \\
       \hline
   \end{tabular}
   \end{center}
   \par\medskip\par
   (3): Flag ``C'' denotes the source was used as 
   an amplitude calibrator; (4): The number of observations;
   (5): Correlated flux density in Jy; (6): uncertainty of
   correlated flux density in Jy.

   \par\smallskip\par

   Table \ref{t:flux} is published in its entirety in
   the electronic edition of the journal. A portion is 
   shown here for guidance regarding its form and content.
\end{table}

\begin{figure}[h]
   \ifpasp
       \includegraphics[width=0.46\textwidth]{fig5.eps}
     \else
       \includegraphics[width=0.47\textwidth]{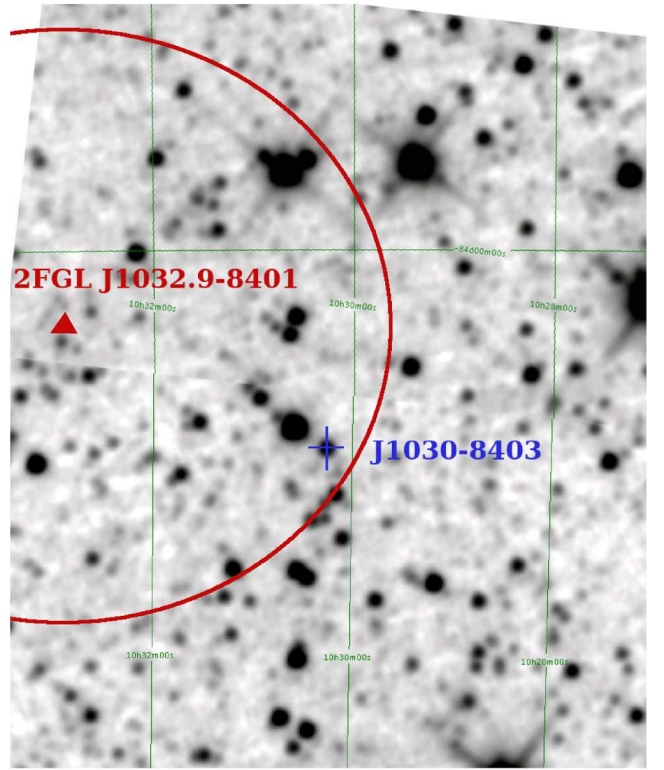}
   \fi
   \caption{Field around {\sf 1031-837} at 4.6~$\mu$m from WISE.
            The 2-$\sigma$ error ellipse around {\sf 2FGL J1032.9-8401}
            position denoted as a triangle is shown with red colour. 
            The blue dot shows {\sf 1031-837} position.
           }
   \label{f:1031-837}
\end{figure}

  Using empirical gain factors, we evaluated the average flux density
of each of the 238 sources observed in this experiment. The average flux 
density of \mbox{\sf 1031-837} was found to be $46 \pm 7$~mJy. Since it was 
detected only in four scans on the \hobart/\wark\ baseline, no information 
about its structure can be derived. Nevertheless, using information that 
we derived from analysis of this experiment, namely, its correlated flux 
density at projected baseline 50~$M\lambda$ and its position, 1.96 \Fermi\ 
position uncertainty, we were able to compute the likelihood ratio of 
its association with the $\gamma$-ray object as defined in 
\citet{r:aofus2}: 22.8. That means that the probability of the 
$\gamma$-ray source and the radio source being the same object is by 
a factor of 22.8 greater than the probability of two unrelated objects 
being incidentally projected close to each other. Thus, we conclude 
that we have established an association between the radio and 
$\gamma$-ray sources.

  There is a source {\sf WISE J103015.28-840308.6} from the Wide-field 
Infrared Survey Explorer (WISE) catalogue (ALLWISE, November 13, 2013 
\citep{r:wise1,r:wise2}) within 90~mas of \mbox{\sf 1031-837} 
position (see Figure~\ref{f:1031-837}). Cross-matching the WISE and the 
Radio Fundamental Catalogue finds 5984 matches. Fitting parameters of 
the Rayleigh distribution to distances between source positions of matches 
we found that the estimate of the 1$\sigma$ uncertainty along each coordinate 
axis is 95~mas. We interpret this estimate as a measure of position accuracy 
of the ALLWISE catalogue for point-like sources. Therefore, we conclude that 
{\sf 1031-837} and {\sf J103015.28-840308.6} are the same object. It was 
detected in all 4~bands of WISE instrument with magnitudes of 14.10, 
13.11, 10.27, and 8.16 at 3.4, 4.6, 12 and 22 $\mu$m, respectively. 
Figure~\ref{f:1031-837} demonstrates that the number of IR sources
within the \Fermi\ position error ellipse is so high, that their 
association with the $\gamma$-ray object is impossible without use of 
additional information.

  Analyzing the flux densities of other sources, we found that the 
typical detection limit for 90~s integration on the  \hobart/\wark\ 
baseline ranged from 40 to 70~mJy with a median value of 50~mJy. Fixing 
the cryogenic problem experienced at \hobart\ promises to reduce 
the detection limit to a 20~mJy level.

\section{Summary and Future Work}      \label{s:concl}

  A new radio astronomy antenna \wark\ has observed successfully its
first scientific VLBI experiment. Its position with uncertainty over
the vertical coordinate 5~cm and horizontal coordinates 1~cm has
been determined. Error analysis shows that the systematic errors are
a factor of 2 smaller than these uncertainties. The averaged SEFD at 
$45\degr$ elevation angle was 650~Jy.

  The first scientific experiment allowed us to establish an association
of \Fermi\ detected $\gamma$-ray source 2FGL J1032.9-8401 that was
previously considered unassociated. Position of its radio counterpart
was determined with the uncertainties of 4.6~mas over both 
coordinate components and its correlated flux density was found 
to be $54 \pm 5$~mJy. We conclude that \wark\ is ready for scientific 
programs.

\begin{figure}[h]
   \ifpasp
        \includegraphics[width=0.46\textwidth]{fig6.eps}
      \else
        \includegraphics[width=0.46\textwidth]{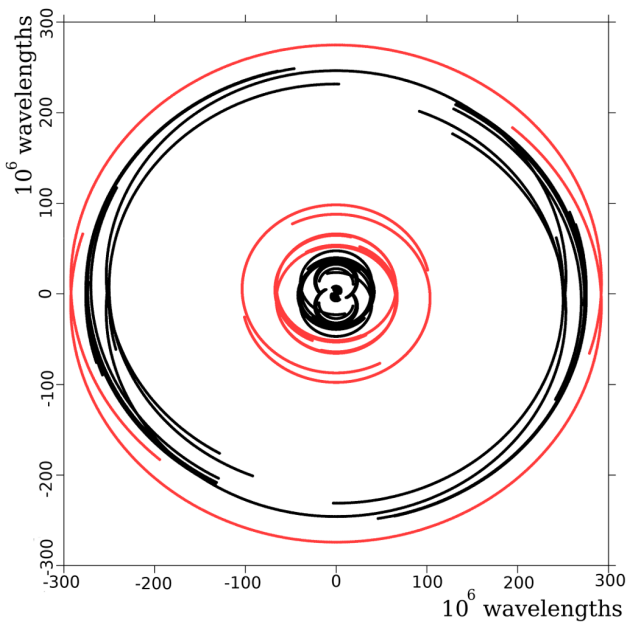}
   \fi
   \caption{The {\it uv} coverage of simulated 12 hour long observations 
            of 1934-638 at a 7-station network 
            {\sc wark30m, ceduna, parkes, hobart26, atca, mopra}, 
            and {\sc hartrao}. Points at baselines with {\sc wark30m} 
            are shown with red.}
   \label{f:uvp}
\end{figure}

  At the moment, there are several VLBI arrays that are actively observing:
the Very Long Baseline Array (VLBA), the European VLBI Network (EVN), 
the VLBI Exploration for Radio Astronomy (VERA), the Korean VLBI Network
(KVN), the Chinese VLBI Network (CVN), the Russian Quasar network, and
the Long Baseline Array. All these arrays, except the last, are located 
in the northern hemisphere and cannot observe sources with declinations
less than $-40\degr$. \wark\ will be a valuable addition to the network
of sensitive southern telescopes (SEFD less than 1000~Jy): 26m Hartrao, 
30m Ceduna, 64m Parkes, 22m Mopra, 6x22m ATCA, and 26~m Hobart. Figure 
\ref{f:uvp} shows the {\it uv}-coverage when this seven-station array
observes the 1934-638. Tracks of baselines with stations to \wark\ are 
highlighted in with red. The main focus of VLBI programs with \wark\ 
is anticipated to be participating in surveys targeting sources in the 
declination zone $[-90\degr, -40\degr]$ and exploring peculiar objects, 
mainly in this declination zone inaccessible to the arrays in the northern 
hemisphere. These programs will be complement to programs that are conducted 
in the northern hemisphere and will convert them into all-sky surveys. 
We anticipate that the most valuable contribution  of the southern VLBI array 
with \wark\ participation will be given to those programs that benefit from 
completeness and all-sky coverage. These programs include the absolute 
astrometry survey that is used for space navigation, for a connection 
of VLBI and {\it Gaia} coordinate systems, as a list of phase calibrators 
for phase referencing programs, and for other important applications, such 
as the VLBI survey of a flux-limited sample of compact radio sources; 
for a search for associations of $\gamma$-ray sources. We have plans 
to equip \wark\ with a cooled C-band receiver, an X-band receiver and with 
the a 2~Gbps recording system in 2015.

  Among the 244 observed sources, 175 have been detected in three or more 
observations. We determined their correlated flux densities with errors
15\% (See Table~\ref{t:flux}). This shows the potential of a southern 
hemisphere VLBI array to contribute to programs of flux density monitoring
of active galaxy nuclei associated with $\gamma$-ray sources. The TANAMI 
program of monitoring 84 radio sources associated with $\gamma$-ray objects 
in the southern hemisphere with the Australian Long Baseline Array (LBA) 
\citep{r:tanami} whose goal is to provide a time series of images. 
However, TANAMI observes 25--30 sources per 24 hour experiment at a 
6--10 station network, i.e. requires more than one order of magnitude more 
resources. A program of coarse flux density monitoring of a much wider list 
will be complimentary to TANAMI. By March 2015, there are 270 known 
$\gamma$-ray sources with declinations $< -30^\circ$ associated with the 
radio sources that exhibit emission from milliarsec regions detected with 
VLBI. All these sources can be observed in a 27--30~hour observing session. 
A program of monthly monitoring of these sources has the potential to check 
their variability and derive light curves.

\Note{
Both VLBI and single-dish spectral observations in C-band can potentially include
spectral line observations of OH maser lines at 6.03 and 6.035~GHz and methanol
masers at 6.7 GHz, which are important for the identification and study of
star-forming regions.  For example, 6.7~GHz methanol masers are exclusively
associated with the early stages of high-mass star formation and VLBI observations
can be used to obtain accurate distance estimates through trigonometric parallax.
Single dish observations with the 30m radio telescope in New Zealand can be used to
monitoring of the variability of methanol maser sources found by the Parkes
multi-beam survey~\citep{r:woodburn}.
}

\section*{Acknowledgments} 

This research has made use of the data products from the Wide-field Infrared 
Survey Explorer and the NASA/ IPAC Infrared Science Archive, which is 
operated by the Jet Propulsion Laboratory, California Institute of 
Technology, under contract with the National Aeronautics and Space 
Administration. We thank Dimitri Duev for careful checking numeric tables.

\end{document}